\documentstyle[12pt]{article}
\input epsf

\begin{document}

\title{Dynamics of the formation of an event horizon}


\author{A. A. Shatskiy, A. Yu. Andreev}


\date{(Submitted 17 November 1988)\\
Zh. \'{E}ksp. Teor. Fiz. {\bf 116}, 353--368 (August 1999)}

\maketitle

\begin{abstract}
The radial motion of matter in a centrally symmetric
gravitational field in a comoving reference frame is investigated
for a realistic equation of state of matter.  The dynamics of the
formation of an event horizon are investigated.
\end{abstract}

\renewcommand{\thesection}{\arabic{section}}
\section{INTRODUCTION}

The formation of a black-hole event horizon has attracted a
great deal of attention on the part of physicists for a long
time.  An enormous amount of material has been written on this
subject (see, for example, Ref. \cite{1,2,3,4});
nevertheless, the treatment of this problem within the general
theory of relativity has created more questions than known
solutions.

One of the main questions concerning this problem is still
the reciprocal influence of accreting matter on a black hole. 
The motion of test particles in the field of a black hole has
been considered hitherto for the most part, but they, as we know,
do not exhibit a reciprocal influence, which can be enormous when
a falling particle achieves the speed of light as it crosses the
event horizon.

{\bf 1.}  In this paper we consider the special, but physically
real case of spherically symmetric accretion on a central body
without allowance for rotation.  The following notation is
adopted: the speed of light $c$ and the gravitational constant
$G$ are set equal to unity.  In these units the gravitational
radius for a given mass $M$ is $r_g = 2M$, i.e., the radius of the
event horizon in free space for the same mass concentrated at the
center.

Let us devise a likely model for the evolution of the
system.  We assume that our system is a cooling massive star
having a radius $R_0$ and a gravitational radius $R_{G_0} = 2M(R_0)$,
where $R_{G_0} < R_0$.  The matter comprising this body is initially
at rest (``dust'' with the equation of state $P 
= \alpha\varepsilon$, where $P$ is
the pressure in the matter, $\varepsilon$ 
is the energy density, and $\alpha$ is
a constant).  In the next moment the matter begins to fall
freely.\cite{1)}$^)$  If it is assumed that the gravitational fields are
not excessively strong and that the dust density\cite{2)}$^)$ is fairly
small in the initial moment, a force field with a finite energy
is needed to retain it in the initial moment.  After this field
is removed, the dust leaves the system and ceases to interact
with it after a time of the order of the size of the system,
i.e., after a time much shorter than the time during which the
dust manages to partially settle and the gravitational fields
increase dramatically.  Thus, this model is physically
consistent.

What subsequently happens to the system?  The dust begins to
fall toward the center of the body, increasing its mean density
and the gravitational radius $r_g(r)$ for the mass $M(r)$ at a
certain radius $r$.  If we would neglect the reciprocal influence
of the pressure of the moving matter on the dynamics of the
system and on its gravitational field, then after all of the
matter has unavoidably fallen and the inequality
$$
r_g(r) = 2M(r) \geq r
\eqno{(1)}
$$
holds at one of the points $r$ of the system, an event horizon
would form at that point according to Schwarzchild's solution for
a gravitational field in a vacuum, i.e., the velocity of the
falling matter relative to the $r = {\rm const}$ surfaces would
reach the speed of light (see below).  Is this what actually
happens?  The achievement of the speed of light by the matter
causes a change in the sign of the interval and is therefore an
invariant event, which does not depend on the choice of the
reference frame.

An attempt to solve this problem in a reference frame which
is stationary at infinity leads at once to a contradiction.  In
fact, the analytically exact, nonstationary model, in principle,
cannot be studied.  If, however, a simplification is made and it
is assumed that the system is quasistationary at a certain moment
in time in the range from one radius to a certain radius known to
be large, but still far smaller than the dimensions of the
system, then it can be stated, at the very least, that  the
components $g_{tt}(r)$ and $g_{rr}(t)$ of the metric have singularities
(zeros and poles) in this reference frame.\cite{3)}$^)$ When the
parameters of the system are chosen so that there would be a
region in space where the inequality (1) is sure to be satisfied,
it becomes clear that the metric does not have singularities,
regardless of whether the inequality (1) is satisfied.

This can be shown by assuming that if a singularity appears
at a certain point $r_0$, the component of the metric near it can
be represented in the form
\[
g_{ii}(r) \approx {\rm const}(r - r_0)^{y_i},
\]
where $y_i$ is a certain number.  When such a metric is
substituted into the equations, it is found that they do not have
a  solution for any $y_i \neq 0$.

This apparently indicates that the singularities and thus
the horizon of the rapidly moving matter are eliminated (the
right-hand side of the Einstein equations, which is equal to zero
in a vacuum, becomes singular in the presence of
ultrarelativistic falling matter when the radial component of the
three-velocity tends to unity and the radial and temporal
components of the four-velocity tend to infinity; this is also
the reason for the elimination of  the singularities of the
metric).

However, in reality all this stems from the inapplicability
of the quasistationary approximation in the case of strong
gravitational fields.  It is inapplicable because the passage of
time in the system is highly nonuniform due to the nonuniformity
of the component $g_{tt}(r)$ of the metric.  This causes the
picture, which appears to be stationary far from the center, to
become highly nonstationary to an observer approaching the
symmetry center of the system.

Nevertheless, this does not remove the question posed: do an
horizon and a black hole appear in the real nonstationary case?

{\bf 2.}  An answer to the question posed can be found by
selecting a comoving reference frame.  The problem was solved in
this frame in Ref. 1 (Sec. 103) in the special case of $\alpha=0$ (see
below).  Matter is at rest in the reference frame chosen, and its
motion can be evaluated only from the variation of the
``circumferential'' or photometric distances $r$, which are related
to the center of the system and are defined as the circumferences
of the respective circles around the center: $2\pi r$.  When the
radius $r$ is defined as such, it is convenient to represent the
metric in the form
$$
ds^2 = e^\nu\,dt^2 - e^\lambda\,dR^2 - r^2(d\theta^2 + 
\sin^2\theta\,d\varphi^2).
\eqno{(2)}
$$
Here $R$ is the coordinate of a dust particle in the comoving
reference frame or its index, and $e^\nu$, $e^\lambda$, and $r$ are
functions of $R$ and $t$.  It is noteworthy that at zero
pressure, i.e., when $\alpha=0$, we have $\nu=0$, i.e., the reference
frame is simultaneously synchronous.

To solve the problem posed we write out the Einstein
equation in the comoving reference frame:
$$
r'^2e^{-\lambda}(1 + r\nu '/r') - e^{-\nu}(2r\ddot{r} 
+ \dot{r}^2 - r\dot{r}\dot{\nu}) =
1 + 8\pi\alpha r^2\varepsilon,
\eqno{(3a)}
$$
$$
2\dot{\mu}' + \dot{\mu}\mu ' - \dot{\lambda}\mu ' - \nu '\dot{\mu} =0,
\eqno{(3b)}
$$
$$
\left(\lambda + 2\mu + \frac{2}{1+\alpha}\ln \varepsilon\right) = 0,
\eqno{(3c)}
$$
$$
\left(\nu + \frac{2\alpha}{1+\alpha}\ln \varepsilon\right)' = 0.
\eqno{(3d)}
$$
Here $\mu = 2\ln r$, a prime denotes differentiation with respect
to $R$, and a dot denotes differentiation with respect to $t$. 
Equations (3) were derived in Ref. \cite{1} [Eqs. (2), (5),
and (6) of problem 5 in Sec. 100].

It follows from (3d) that
\[
\nu = -\frac{2\alpha}{1+\alpha}\ln \varepsilon + f^*(t)
\]
and that by transforming the time $t$ in the interval element (2)
the function $f^*(t)$ can be set equal to $[2\alpha/(1 
+ \alpha)]\ln \varepsilon_*$,
where $\varepsilon_*$ is a constant with the dimensions of energy density,
which expresses the measurement scale of $\varepsilon$.  Then
$$
\nu = \frac{2\alpha}{1+\alpha}\ln \frac{\varepsilon}{\varepsilon_*}.
\eqno{(4)}
$$
We next assign the indices $R$ to the dust particles so that
$r=R$ in the initial moment.  Under such initial conditions
$r'(R, t)$ corresponds to $(n_0/n)^{1/3}$, where $n(R,t)$ is the
concentration of dust particles and $n_0$ is its value at the
initial moment.

Let us now ascertain the conditions which must be imposed on
the initial distribution of the dust.  The most important among
them is that the inequality (1) need not hold within the matter
at the initial moment.  It means that there is no horizon in all
space in the initial moment.  It thus imposes an upper limit on
the initial density of the dust and on the initial dimensions of
the system.  More specifically, if the initial density
distribution of the dust is set equal to $\varepsilon_0(R)$, then, according
to (1), the maximum radius of the body $R_{\max}$ is uniquely
specified by the expression
$$
R_{\max} = 2 \int\limits_0^{R_{\max}} 4\pi\varepsilon_0(R)R^2\, dR.
\eqno{(5)}
$$
Then, it follows from (3c) and (4) that
\[
\frac{\partial}{\partial t}[\alpha(\lambda + 2\mu) - \nu] = 0
\]
or
$$
\nu = \alpha[\lambda + 2\mu + f^*(R)],
\eqno{(6)}
$$
where $f^*(R)$ is an arbitrary function that depends on the
initial conditions.

{\bf 3.}  Let us now find the initial values for all the variables
in our problem.  We have already assigned these values for $r$
and $\varepsilon$.  From (4) it follows that 
$$
\nu_0 = \frac{2\alpha}{1+\alpha}\ln \frac{\varepsilon_0}{\varepsilon_*}.
\eqno{(7)}
$$
To find the initial value of $\lambda$ we take advantage of the fact
that the problem has already been solved for $\alpha=0$, and we can
therefore utilize the familiar expression for $\lambda_0|_{\alpha=0}$ from Ref.
\cite{1} (Sec. 103.6):
$$
\lambda_0(R) = -\ln [1 - S(R)],
\eqno{(8)}
$$
where for $\alpha=0$ we have
$$
S(R) = 2M(R)/R,
\eqno{(9)}
$$
and $M(R)$ is the mass within the radius $R$ at the initial
moment.

The expression for $S(R)$ for an arbitrary value of $\alpha$ is
the same.  It can be obtained from Eq. (4) in problem 5 of Sec.
100 in Ref. \cite{1}, where the Einstein equations in
matter in the comoving reference frame were found for a centrally
symmetric system.  We write out this equation: 
$$
-e^{-\lambda}\left[\mu '' + 
\frac{3}{4}\mu '^2 - \frac{\mu '\lambda '}{2}\right] +
\frac{1}{r^2} + \frac{1}{2e^\nu}\left[\dot{\lambda}\dot{\mu} + (\dot{\mu})^2/2
\right] = 8\pi\varepsilon.
\eqno{(10)}
$$
Expressing $\mu$ in terms of $r$ ($\mu = \ln r^2$) and combining
similar terms, we can bring this expression into the form
$$
8\pi r'\varepsilon r^2 = -[r(r'^2e^{-\lambda} - 1)]' 
+ \frac{r'}{e^\nu}\left[\dot{\lambda}r\dot{r} +
(\dot{r})^2\right].
\eqno{(11)}
$$
Taking into account the expression (8), as well as the fact that,
according to the expression (100.23) in Ref. \cite{1}, the
equality
\[
2M(r) = \int\limits_0^r 8\pi\varepsilon(\tilde{r}, t)\tilde{r}^2 \,
d\tilde{r}|_{t={\rm const}}
\]
holds for the initial moment in time, when $\dot{r}=0$ and
$r'=1$, we obtain the expression (9) for $S(R)$ after
preliminarily integrating (11) over $R$ from 0 to $R$.

Substituting the expression (8) into (6) and taking into
account (7), we find that 
$$
f^*(R) = -\frac{2}{1+\alpha}\ln\frac{\varepsilon_0}{\varepsilon_*} 
+ \ln [1 - S(R)] - \ln R^4.
\eqno{(12)}
$$

{\bf 4.}  Now, plugging (6) into (3b) and dividing everything by
$\dot{\mu}\mu '$, we obtain the expression
$$
\frac{1}{\dot{r}}(2\ln \mu ' + \mu - \lambda)^\cdot = \frac{\nu '}{r'} 
= \alpha\frac{[\lambda +
2\mu + f^*(R)]'}{r'}.
\eqno{(13)}
$$
Taking into account that $e^{-\nu}(2r\ddot{r} + \dot{r}^2 -
r\dot{r}\dot{\nu}) = (e^{-\nu }r\dot{r}^2)^\cdot/\dot{r}$ and introducing the
notation 
$$
U(R, t) = (\dot{r})^2,\ \ Q(R, T) = r'^2e^{-\lambda},
\eqno{(14)}
$$
we see that Eq. (3a) can be written as an equation for $U$:
$$
\frac{\dot{U}}{\dot{r}} + aU = \sigma,
\eqno{(15)}
$$
where
\[
a(R, t) = \frac{1}{r}\left(1 - \frac{r\dot{\nu}}{\dot{r}}\right),\
\ \sigma(R, t) = \frac{1}{r}\left[Q\left(1 + \frac{r\nu '}{r'}\right) - 1
- 8\pi\alpha r^2\varepsilon\right]e^\nu.
\]

This equation has a solution which satisfies the initial
conditions:
$$
U(R, t) = \frac{1}{\gamma^*(R,t)}\int\limits_0^t 
\gamma^*(R, \tilde{t})\sigma(R,
\tilde{t}) \tilde{r}\, d\tilde{t},\ \ \gamma^*(R, t) = \exp
\left[\int\limits_0^t a(R, \tilde{t}) \tilde{r}\,
d\tilde{t}\right].
\eqno{(16)}
$$
Finding $U$, we can obtain an expression for the square of the
velocity of the matter relative to the $r = {\rm const}$ surfaces
from the form of the metric (2) (see Appendix 1):
$$
V^2(R, t) = Ue^{-\nu }e^\lambda/r'^2.
\eqno{(17)}
$$
The expression for $\gamma^*$ can easily be found:
\[
\gamma^*(R, t) = C(R)re^{-\nu},
\]
where the multiplier $C(R)$ for $\gamma^*(R, t)$, which does not depend
on $t$, can be taken out of the integral sign in (16) and
canceled; therefore, it can be set equal to unity.  Then
\[
\gamma^*\sigma = r'^2e^{-\lambda}\left(1 + 
\frac{r\nu '}{r'}\right) - 1 - 8\pi\alpha r^2\varepsilon.
\]
Alternatively, taking into account that Eq. (13) can now be
rewritten as
$$
\frac{(2\ln \mu ' + \mu - \lambda)^\cdot}{\dot{r}} 
= \frac{(\ln Q)^\cdot}{\dot{r}} =
\frac{\nu '}{r'},
\eqno{(18)}
$$
we obtain
$$
\gamma^*\sigma = \frac{(r(Q-1))^\cdot}{\dot{r}} - 8\pi\alpha r^2\varepsilon.
\eqno{(19)}
$$
Then (16) is rewritten in the form
$$
U = \frac{e^\nu}{r}[r(Q - 1) - R(Q_0 - 1) + 2\alpha m(R, t)],
\eqno{(20)}
$$
where we have introduced the notation
$$
m(R, t) = \int\limits_t^0 4\pi\tilde{\varepsilon}\tilde{r}^2\tilde{r}\,
d\tilde{t} = \int\limits_r^R 4\pi\tilde{\varepsilon}\tilde{r}^2\tilde{r}\,
d\tilde{r}|_{R={\rm cosnt}}.
\eqno{(21)}
$$

{\bf 5.}  For $\alpha = 0$, taking into account (6), (9), (18), and
(20), we can easily obtain an analytically exact expression for
$U$ and $V$:
$$
U_{\alpha=0} = S(R)\left(\frac{R}{r} - 1\right) = \frac{2M(R)}{r} - S(R).
\eqno{(22)}
$$
Substituting this expression into (17), for the velocity we
obtain
$$
V_{\alpha=0}^2 = \frac{2M(R)/r - S(R)}{1 - S(R)}.
\eqno{(23)}
$$
Hence $V_{\alpha=0} = 1$ when $r = r_0 = 2M(R)$.  This coincides with the
results in Sec. 100 of Ref. \cite{1}, where the problem has
already been solved for this case.

{\bf 6.}  Finally, let us consider the location of the
horizon\cite{4)}$^)$ in the presence of a nonzero pressure.  For this
purpose we plug the expressions (20) and (8) for $e^{-\lambda_0}$ into
formula (17).  After some relatively simple transformations, we
ultimately obtain
$$
r = \frac{2M(R) + 2\alpha m(R, t)}{1 - Q(1 - V^2)}.
\eqno{(24)}
$$
As will be shown in Appendix 1, the horizon appears at the point
and at the time where the velocity of the falling matter relative
to the $r = {\rm const}$ surfaces reaches unity, i.e., where $V =
1$.  In addition, the speed of light relative to the falling
matter at this site is also, as always, equal to unity.

Hence, according to (24) and (9), the horizon radius $r_{\rm
hor}$ is given by the formula
$$
r_{\rm hor} = 2M(R) + 2\alpha m(R, t_{\rm hor}).
\eqno{(25)}
$$
Thus, the horizon is displaced to a larger radius in comparison
to the value in a vacuum $r_{0_{\rm hor}} 
= 2M(R)$ by $2\alpha m(R, t_{\rm hor})$. 
In this case the quantity $m(R, t)$ has the meaning of the mass
which would accumulate if we would join layers of dust with the
initial radius $R$ and the thickness $d\tilde{r}(\tilde{t})$ to
one another up to the radius $r(R, t)$ at the moment when this
$d\tilde{r}$ layer passes through the joining point.

{\bf 7.}  Regarding the possible values of $\alpha$ we note that $\alpha=0$
corresponds to dustlike matter without interactions between the
particles.  The results obtained for them are the same [see (23)]
as the results for test particles in a central field of mass $M$
(see Sec. 101 in Ref. \cite{1}).  However, of course, such
an equation of state of matter cannot correspond to reality near
the horizon.  It is reasonable to assume that the
ultrarelativistic equation of state of matter, in which $\alpha =
1/3$, holds near the horizon.  Therefore, the location of the
horizon should probably be sought with just such a value of $\alpha$.

{\bf 8.}   When $\alpha \neq 0$, it would appear that the falling matter
should be slowed under the action of the pressure gradient, and
the horizon should therefore form later, i.e., be displaced
toward smaller values of $r$, but, as we have just shown, it is
displaced toward larger values of $r$ by $2\alpha m(R, t_{\rm hor})$.  What
is the reason for this contradiction?  It can be seen from the
initial equations (3) that the reason should be sought in Eq.
(3a).  For this purpose we explore Eqs. (3a) and (4) in the
initial moment for the case of $\alpha \ll 1$.  In that moment $\dot{r}
= 0$ and $r' = 1$; therefore, we write
\[
[1 - S(R)]\left(1 -2ar\frac{\varepsilon '}{\varepsilon}\right) -
2r\tilde{r}\left(\frac{\varepsilon}
{\varepsilon_*}\right)^{2\alpha} \approx 1 + 8\pi\alpha r^2\varepsilon.
\]
Since
\[
(\varepsilon/\varepsilon_*)^{2\alpha} \approx 1 
+ 2\alpha \ln \frac{\varepsilon}{\varepsilon_*},\ \ S(R) =
\frac{2M(R)}{R},\ \ r = R,
\]
then, after performing some relatively simple transformations, in
the linear approximation with respect to $\alpha$ we obtain
$$
\tilde{r} = -\frac{GM(R)}{r^2}\left[1 - 
2\alpha\ln \frac{\varepsilon}{\varepsilon_*}\right]
- \frac{\nabla P}{\rho}\left[1 - 
\frac{2GM(r)}{rc^2}\right] - 4\pi\alpha Gr\rho,
\eqno{(26)}
$$
where $\rho(R, r) = \varepsilon(R, r)/c^2$ is the density of the matter.  Here,
for the sake of clarity we use the ordinary (Gaussian) system of
units with $G \neq 1$ and $c \neq 1$.  It can be seen from (26) that
the first term corresponds to the ordinary Newtonian force of
gravity, and the second term corresponds to the interaction force
between the particles, i.e, the pressure gradient (just this
force is the cause of the slowing of the fall of the matter in
the first stage).  The remaining terms do not appear in the
equation of motion in the Newtonian approximation (the
corrections in square brackets are also neglected in that case),
but, as we have already seen, the last term begins to dominate
over the second term at high energies; therefore, a shift of the
horizon toward larger radii appears.  Thus, the contradiction has
been resolved.  Physically this corresponds to the ``gravity of
pressure'' in the general theory of relativity, which surpasses
the gradient terms at high energies.

{\bf 9.} The analysis performed allows us to draw the following
conclusions.

First, a shift of the horizon toward a larger radius in
comparison to the Schwarzchild radius due to the ``gravity of
pressure'' has been discovered.  We stress that this effect is
purely dynamic and is not observed in the static case (after all
the matter has fallen).

Second, according to the results in Appendix 2, the
evolution of the entire system at a constant value of $\alpha$ is
completely specified by the energy density distribution profile
in the initial moment, i.e., for example, by the normalized
density distribution of the matter and by the value of the
parameter $S$ at an arbitrary point on this distribution.

If the evolution of only one spherical layer of matter with
the index $R$ must be described, it is completely specified by
three dimensionless parameters in the initial moment in that
layer and, in this sense, does not depend on the initial
distribution of the matter in the system below and above that
layer.  However, this in no way signifies the independence of the
spherical layers in the general case, since just these three
parameters, as will be seen from Appendix 2, govern the
interaction of the layers.  Consequently, integration of the
system leads to a complete family of self-similar solutions.

Third, according to Appendix 2, a local extremum appears on
the $V(R)|_{t={\rm const}}$ curve for a specific choice of initial
parameters, and when $V = 1$, it leads to the formation of a
second apparent horizon in the system (an analog of the second
horizon in the Reissner--Nordstr\"{o}m and Kerr--Newman solutions for
an electrically charged rotating static black hole; for an
interpretation of these solutions, see, for example, Refs.
\cite{5} and \cite{6}).

\appendix
\section*{1}

We have hitherto used the term horizon to refer to a
trapping surface, or an apparent horizon, as it is called in the
literature.

Let us ascertain the difference between an event horizon and
an apparent horizon in greater detail in an example.  We assume
that we already have a stationary black hole of mass $M$ and that
there is an apparent horizon at $r = 2M$.  Now we assume that
another chunk of matter with a mass $\delta M$ falls into our black
hole.  After it falls, the radius of the apparent horizon
increases to $2(M + \delta M)$.  Thus, if an observer is placed between
these radii before the additional chunk of matter falls, he would
then be outside the black hole, but after the chunk of matter
falls he would be inside it.  The concept of an event horizon is
global and is determined by the entire course of evolution of the
black hole or, stated differently, by all the mass which falls
into it at any time.

The existence of an apparent horizon, which specifies a
black hole locally, is sufficient for the existence of a black
hole.  As follows from our arguments, in the spherically
symmetric case the two horizons ultimately coincide and form a
static black hole described by Schwarzchild's solution. 
Therefore, we shall henceforth use the term horizon to refer to
the apparent horizon.

Let us prove that the horizon in a system with spherical
symmetry forms at the moment when a falling particle with a
nonzero rest mass achieves the speed of light relative to the $r
= {\rm const}$ surfaces at the same point.  For this purpose we
write the law of motion for the particle in the form
$$
r(R, T) = R - \int\limits_0^t \sqrt{U(R, \tilde{t})}\, d\tilde{t}.
\eqno{(27)}
$$
We now assume that we are located on a dust particle with the
index $R_\infty$ and we are tracking a dust particle with the index
$R_p$, which sends us a light beam passing through the radii
$r_p(R_p, t)$, from the large radius $r_\infty(t)$.  The criterion for
determining that the dust particle has not yet reached the
horizon is the fact that we still see light from it, i.e., the
light propagating still crosses the radii $r > r_p$.  Therefore,
the criterion for determining that the dust particle has reached
the horizon is an event in which the light propagating from $R_p$
can no longer cross the radii $r > r_p$.  Let us express this
criterion mathematically.

\begin{figure}[t]
\centering
\epsfbox[150 430 400 760]{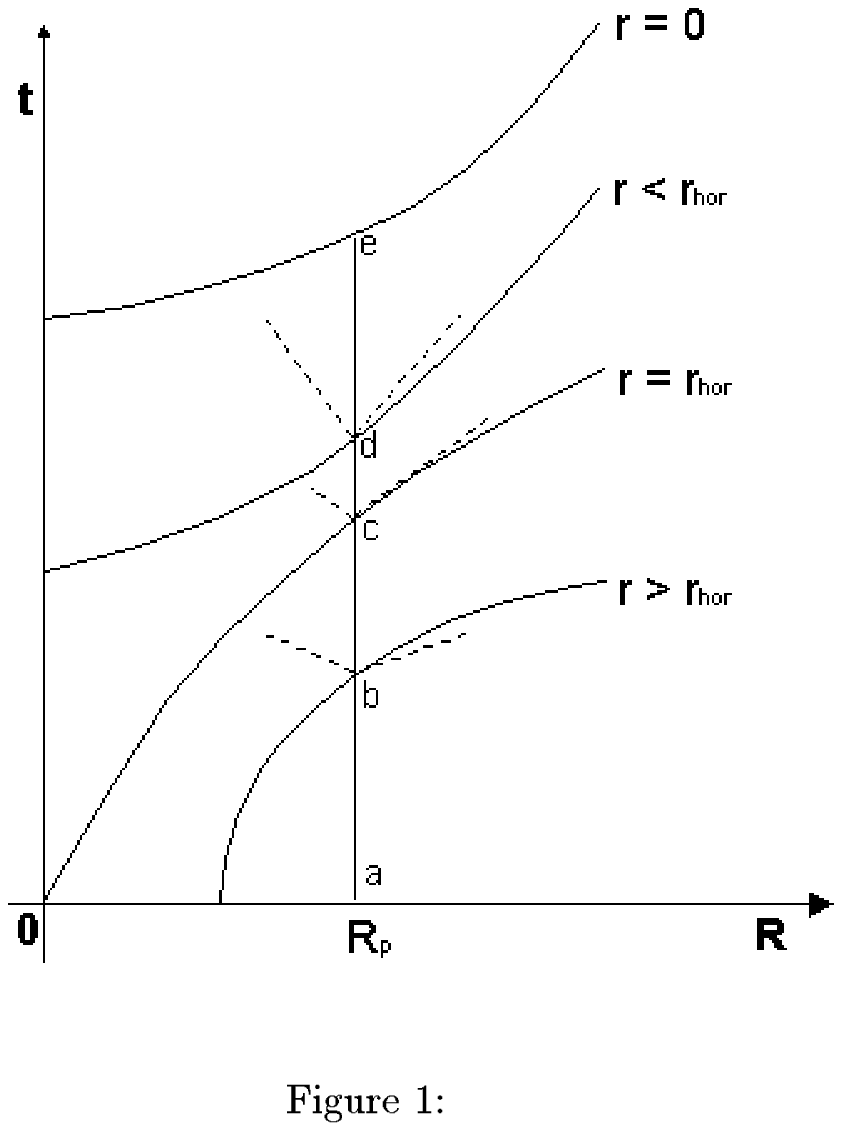}
\label{1}
\end{figure}

In Fig. \ref{1} the vertical straight line $abcde$ denotes
the world line of an $R_p$ dust particle in the coordinates $R$
and $t$ of the comoving reference frame from the moment of rest
($a$) to the center of the system ($e$) at $r=0$.  In this case
of solid curves passing through points $e$, $d$, $c$, and $b$
denote, respectively, lines of constant values of $r(R, t)$ for
$r=0$, $r < r_{\rm hor}$, $r = r_{\rm hor}$, and $r > r_{\rm hor}$.  The dashed
lines emerging from these points denote the cones within which
light emitted by the $R_p$ dust particle can propagate (light
cannot propagate outside these cones).  Therefore, according to
the criterion indicated above, the horizon forms at the point
where the cone is tangent to the $r = {\rm const}$ line.  In the
figure this line is designated as $r = r_{\rm hor}$, and it passes
through point $c$.  For clarity, Fig. \ref{1} shows that the
light cone intersects lines with $r > r_p$ at point $b$;
therefore, there is still no horizon at that point.  This figure
also shows that at point $d$ the light cone is located entirely
above the $r = {\rm const}$ curve passing through point $d$. 
Consequently, this light cone intersects only lines with $r <
r_p$, and therefore point $d$ is already located below the
horizon.

Let us examine the expression (27) on one of the $r = {\rm
const}$ curves and take its complete differential on that curve:
\[
0 = dR - \sqrt{U}\, dt - \frac{1}{2}\int\limits_0^t \frac{U'(R,
\tilde{t})}{\sqrt{U(R, \tilde{t})}}\,dR\, d\tilde{t},
\]
or
$$
\sqrt{U}\left.\frac{dt}{dR}\right|_{r={\rm const}} = 1 -
\frac{1}{2}\int\limits_0^t \frac{U'(R, \tilde{t})}{\sqrt{U(R,
\tilde{t})}}\, d\tilde{t}.
\eqno{(28)}
$$
Next, differentiating (27) with respect to $R$, we obtain the
following expression for $r'$:
\[
r' = 1 - \frac{1}{2}\int\limits_0^t
\frac{U'}{\sqrt{U}}\,d\tilde{t},
\]
with consideration of which from (28) we find
$$
\left.\frac{dt}{dR}\right|_{r={\rm const}} = \frac{r'}{\sqrt{U}}.
\eqno{(29)}
$$
Thus, we have found an expression for the slope of an $r = {\rm
const}$ curve relative to the $R$ axis.

To find the slope of a light cone, by definition, for light
we have $ds^2 = 0$.  Hence, from (2) it follows that
$$
\left.\frac{dt}{dR}\right|_{\rm ligth} = \sqrt{e^{\lambda-\nu}}.
\eqno{(30)}
$$
According to the foregoing statements, the criterion for the
absence of a horizon is the condition
$$
\left.\frac{dt}{dR}\right|_{\rm ligth} < \left.\frac{dt}{dR}\right|_{r={\rm
const}}.
\eqno{(31)}
$$
Substituting the expressions (29) and (30) therein and taking
into account (17), we obtain this criterion in the form
$$
|V| = \frac{\sqrt{Ue^{\lambda-\nu}}}{r'} < 1.
\eqno{(32)}
$$
Here, according to (29), the rate of motion of the matter
relative to the $r = {\rm const}$ lines has the form
\[
|V| = \left.\frac{dl}{d\tau}\right|_{r={\rm const}} =
\sqrt{e^{\lambda-\nu}}\left.\frac{dt}{dR}\right|_{r={\rm const}}.
\]

Thus, the assertion that a horizon forms at the moment when
the matter achieves the velocity $V=1$ relative to the $r = {\rm
const}$ surfaces has been proved.  The horizon surface separates
regions in which $r$ is space-similar and time-similar.

\section*{2}

To solve the equations describing collapse, we first bring
them into dimensionless forms.  For this purpose it is convenient
to introduce the following notation:
\[
x = r/R,\ \ \gamma = 
\frac{\rho_0(R)}{\langle\rho\rangle} = \frac{8\pi\varepsilon_0(R)R^2}{3S(R)}.
\]

In this Appendix we find the ranges of permissible values of
$\gamma$ and $S$, investigate the character of collapse at these
values of the parameters, and obtain numerical solutions for
$V^2$.  We must first of all know the form of the function
$r'(x)$.  Differentiating (27), we obtain\cite{5)}$^)$
$$
r'(x) = 1 + \frac{1}{2}R \int\limits_1^x [\ln U(R, \tilde{x})]'\,
d\tilde{x}.
\eqno{(33)}
$$
Unfortunately, an analytically exact expression for $r'$ can be
found only in the case of $\alpha=0$, the character of collapse can be
assessed exactly only at that value of $\alpha$.  However, the main
features of that character, as will be seen below, remain the
same as in the case of $\alpha\neq0$.  Therefore, let us first
investigate the case of $\alpha=0$.

Thus, we should find $r'(R,x)$.  According to the expression
(22) for $U$, we obtain
\[
\ln [U(R, x)] = \ln [S(R)] + \ln \left(\frac{1}{x}-1\right).
\]
Introducing the notation $y = r' - x$ and taking into account
that $x' = y/R$, we have
\[
(\ln U)' = \frac{S'}{S} - \frac{y}{Rx(1-x)}.
\]
The substitution of this expression into (33) gives
$$
y(x) + x - 1 = \frac{1}{2}\int\limits_1^x \left[\frac{RS'}{S} -
\frac{y(\tilde{x})}{\tilde{x}(1-\tilde{x})}\right]\, d\tilde{x}.
\eqno{(34)}
$$
Differentiating (34) with respect to $x$, we obtain
$$
\frac{\partial y(x)}{\partial x} + a^*(x)y(x) = \sigma^*(R),
\eqno{(35)}
$$
where we have introduced the notation
\[
a^*(x) = \frac{1}{2x(a-x)},\ \ \sigma^*(R) = \frac{RS'}{2S} - 1 =
-\frac{3}{2} + \frac{3}{2}\gamma.
\]
As can be seen, Eq. (35) coincides in form with Eq. (15), and the
initial conditions, $y|_{t=0} = 0$, are the same; therefore, the
method used to solve it is similar.  The solution has the form
$$
y(x) = r' - x = \sigma^*\left[\sqrt{\frac{1-x}{x}} \arctan
\sqrt{\frac{1-x}{x}} - (1 - x)\right].
\eqno{(36)}
$$

Let us find the domain of $r'$.  First, the condition for
compression of the matter has the form $r' \leq 1$.  Second, the
condition that dust layers with different $R$ do not
intersect\cite{6)}$^)$ has the form $r' > 0$.  Thus,
$$
0 < r' \leq 1.
\eqno{(37)}
$$

We assume that the $V^2(R)$ curve for $t = t_m = {\rm const}$
has a local extremum, and we presume (to fix ideas) that it is a
maximum.  Then the horizon appears specifically at the local
maximum, i.e., the point $R = R_{\rm extr}$.  We now find the
condition for a maximum.  First, at that point we should have
$V^2(R_{\rm extr}, x) = V_{\rm extr}^2$.  Second, since it is the first point
at which the velocity of the matter achieves the value $V_{\rm extr}$
and the rate of collapse increases with time, in the vicinity of
this point we should have $V^2 < V_{\rm extr}^2$, or
\[
\frac{\partial V^2(R, t_m)}{\partial R} > 0,\ \ R < R_{\rm extr},
\]
$$
\eqno{(38)}
$$
\[
\frac{\partial V^2(R, t_m)}{\partial R} < 0,\ \ R > R_{\rm extr}.
\]

If it turns out that (38) holds with opposite inequality
signs, there will be a local minimum on the $V^2(R)$ curve at the
point $R_{\rm extr}$ at the moment when the velocity $V_{\rm extr}$ is
achieved at that point, i.e., the matter will achieve the
velocity $V_{\rm extr}$ last at that point.

The condition for an extremum is written in the form
\[
\frac{\partial V^2(R, x)}{\partial R} = 0,
\]
where, according to (23),
\[
V^2(R,x)|_{\alpha=0} = \frac{1-1/x}{1 - a/S(R)}.
\]
Differentiating this expression with respect to $R$, we obtain
$$
\left.\frac{\partial V^2}{\partial R}\right|_{t=t_m} 
= \frac{S/R}{1-S}\left[-y +
\frac{1-3\gamma}{1-S}\right],
\eqno{(39)}
$$
where it has been taken into account that $x' = y/R$ and $S'/S =
(3\gamma-1)/R$.  Then, with allowance for the fact that $0 < x \leq
1$, $0 < \gamma \leq 1$, $-1 < y \leq 0$, and $0 < S < 1$, the
condition (38) can be rewritten in the form
\[
-y > \frac{3\gamma-1}{1-S},\ \ R < R_{\rm extr},
\]
$$
\eqno{(40)}
$$
\[
-y < \frac{3\gamma-1}{1-S},\ \ R > R_{\rm extr}.
\]
If we introduce the notation $z = \sqrt{(1-x)/x}$ and take into
account that, according to formula (23), $z = V\sqrt{1/S - 1}$,
from (40) we obtain 
$$
{3\over 2}(1-\gamma)\left[z\arctan(z) - {z^2\over 1+z^2}\right] -
{3\gamma-1\over 1-S} > 0,\ \ R < R_{\rm extr},
$$
\[
\frac{3}{2}(1-\gamma)\left[z\arctan(z) - \frac{z^2}{1+z^2}\right] -
\frac{3\gamma-1}{1-S} < 0,\ \ R > R_{\rm extr},
\]
or for the extremum point we can write
\[
(1-\gamma)\left[z\arctan(z) - \frac{z^2}{1+z^2} + \frac{2}{1-S}\right] -
\frac{4/3}{1-S} = 0.
\]
$$\eqno{(41)}$$

\begin{figure}[t]
\centering
\epsfbox[150 430 400 760]{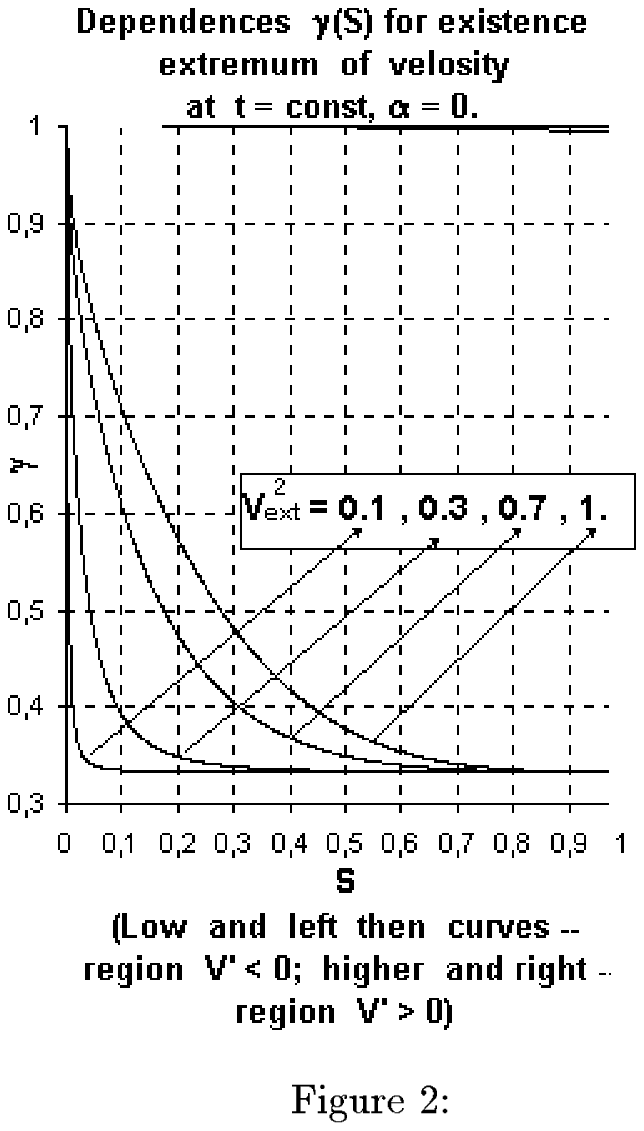}
\label{2}
\end{figure}

This formula can be used to construct the plot of $\gamma(S, V_{\rm extr})$
separating positive and negative values of the derivative $(V^2)'$
and to determine the character of the extremum.  The
corresponding curves for various values of $V_{\rm extr}$ are shown in
Fig. \ref{2}.  The regions where $V' > 0$ are located above and
to the right of them, and the regions where $V' < 0$ are located
below and to the left of them.

It is seen from Fig. \ref{2} that there can be (for a
definite choice of the distribution profile of the matter in the
initial moment and of the parameter $S$ at a certain point $R_*$)
two values of $R$, at which $V = 1$ at a certain moment in time,
and, therefore, the appearance of a second horizon in the system
is possible.

The appearance of a second horizon is not news in the
physics of black holes (see, for example, the Reissner--Nordstr\"{o}m
or Kerr--Newman solution in Ref. \cite{7}).

The results obtained in this Appendix apply to the case of
the absence of pressure, although the case of $\alpha = 1/3$ is of
experimental interest.  Therefore, we used formulas (4), (6),
(8), (9), (12), (17), (18), (20), and (33) to introduce new
dimensionless variables ($\hat{\nu} = 
\nu - \nu_0$, $\hat{\lambda} = \lambda - \lambda_0$,
$\hat{U} = Ue^{-\nu_0} $, and $\hat{\varepsilon} 
= \varepsilon/\varepsilon_0$) and equations for them. 
The initial conditions for them take the form
\[
\hat{\nu}_0 = \hat{\lambda}_0 
= \hat{U}_0 = 0,\ \ r'_0 = \hat{\varepsilon}_0 = 1.
\]
Designating the new coordinates as $x = r/R$ and $\xi = R/R_*$ ($R_*
= {\rm const}$) and introducing the parameters\cite{7)}$^)$ 
\[
h = \xi \partial_\xi \nu_0,\ \ h_{\rm cr} 
= S\frac{1+3\alpha\gamma}{1-S},\ \ \eta = h/h_{\rm cr},
\]
we obtain equations for the new variables in the form
\[
e^{\hat{\nu}/\alpha} = x^4e^{\hat{\lambda}},
\]
\[
\hat{\varepsilon} = (e^{-\hat{\nu}/\alpha})^{(1+\alpha)/2},
\]
\[
\ln (r'^2e^{-\hat{\lambda}}) 
= \int\limits_1^x \left[\xi\frac{\partial_\xi\hat{\nu}}{r'} +
\frac{h}{r'}\right]\, d\tilde{x},
\]
$$
\eqno{(42)}
$$
\[
\hat{U} = e^{\hat{\nu}}\left[r'^2e^{-\hat{\lambda}}(1-S) - 1 +\frac{S}{x} -
\alpha\frac{3\gamma S}{x}\int\limits_1^x 
\hat{\varepsilon}\tilde{x}^2\, d\tilde{x}\right],
\]
\[
r' = 1 + \frac{1}{2}\int\limits_1^x \xi\partial_\xi(\ln \hat{U})\, d\tilde{x} -
\frac{h}{2}(1-x).
\]
In the new variables the velocity is
\[
V^2 = \frac{\hat{U}e^{\hat{\lambda}-\hat{\nu}}}{(r')^2(1-S)}.
\]
Hence $x_{\rm hor} = S + 2\alpha\tilde{m}$, where
$$
\tilde{m} = \frac{m}{R} = 
\varepsilon_0R^2 \int\limits_x^1 4\pi\hat{\varepsilon}\tilde{x}^2\,
d\tilde{x},\ \ \varepsilon_0R^2 = \frac{3\gamma S}{8\pi}.
\eqno{(43)}
$$

We note that this formula and formula (36) can be used to
find the corrections $\delta r_{\rm hor}$ in (25) to the displacement of the
horizon in the linear approximation with respect to $\alpha$, since,
according to Eq. (103.11) from Ref. \cite{1} for $\alpha = 0$,
we have
\[
8\pi\varepsilon r^2 = \frac{2M'}{r'} = \frac{8\pi\varepsilon_0R^2}{r'},
\]
or
\[
\hat{\varepsilon}x^2 = 1/r'.
\]
This expression can be substituted into (43) and a quadrature
expression can be obtained for the correction sought.
\begin{figure}[t]
\centering
\epsfbox[150 436 400 750]{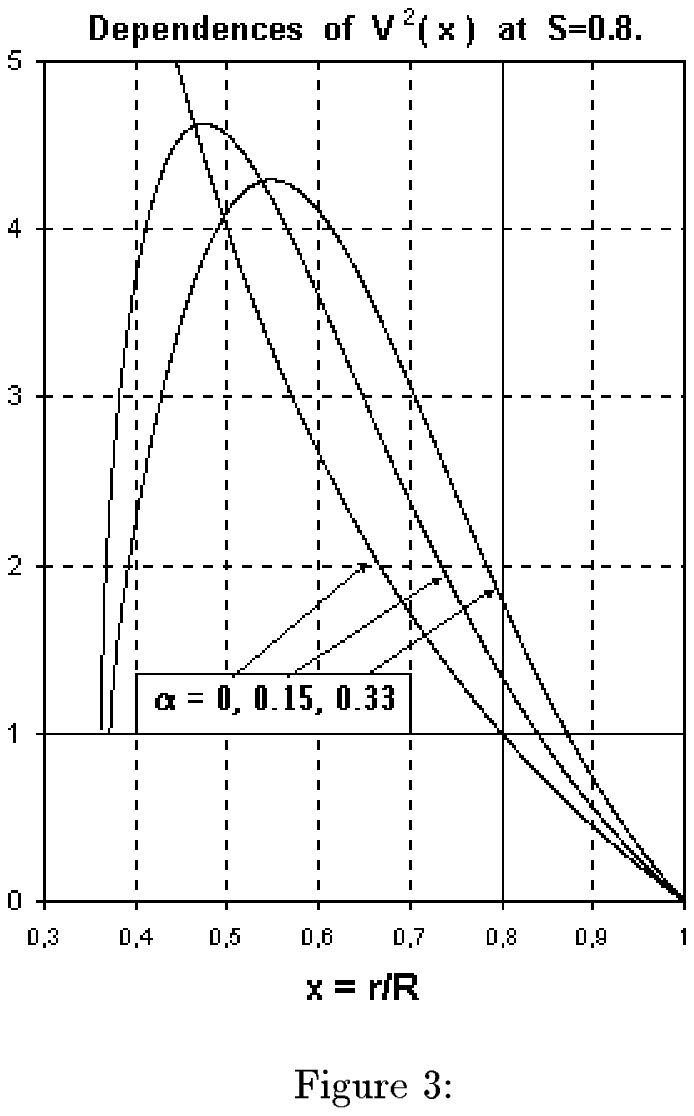}
\label{3}
\end{figure}

In addition, we numerically integrated the equations for the
case of $\alpha\neq0$ using a difference scheme, and the results for
various values of $\alpha$ are presented in Fig. \ref{3}.  As it
should be, according to (25), the plots of $V^2(x)$ are displaced
upward and to the right as $\alpha$ is increased from $\alpha = 0$ to $\alpha =
1/3$.

\begin{figure}[t]
\centering
\epsfbox[79 440 520 760]{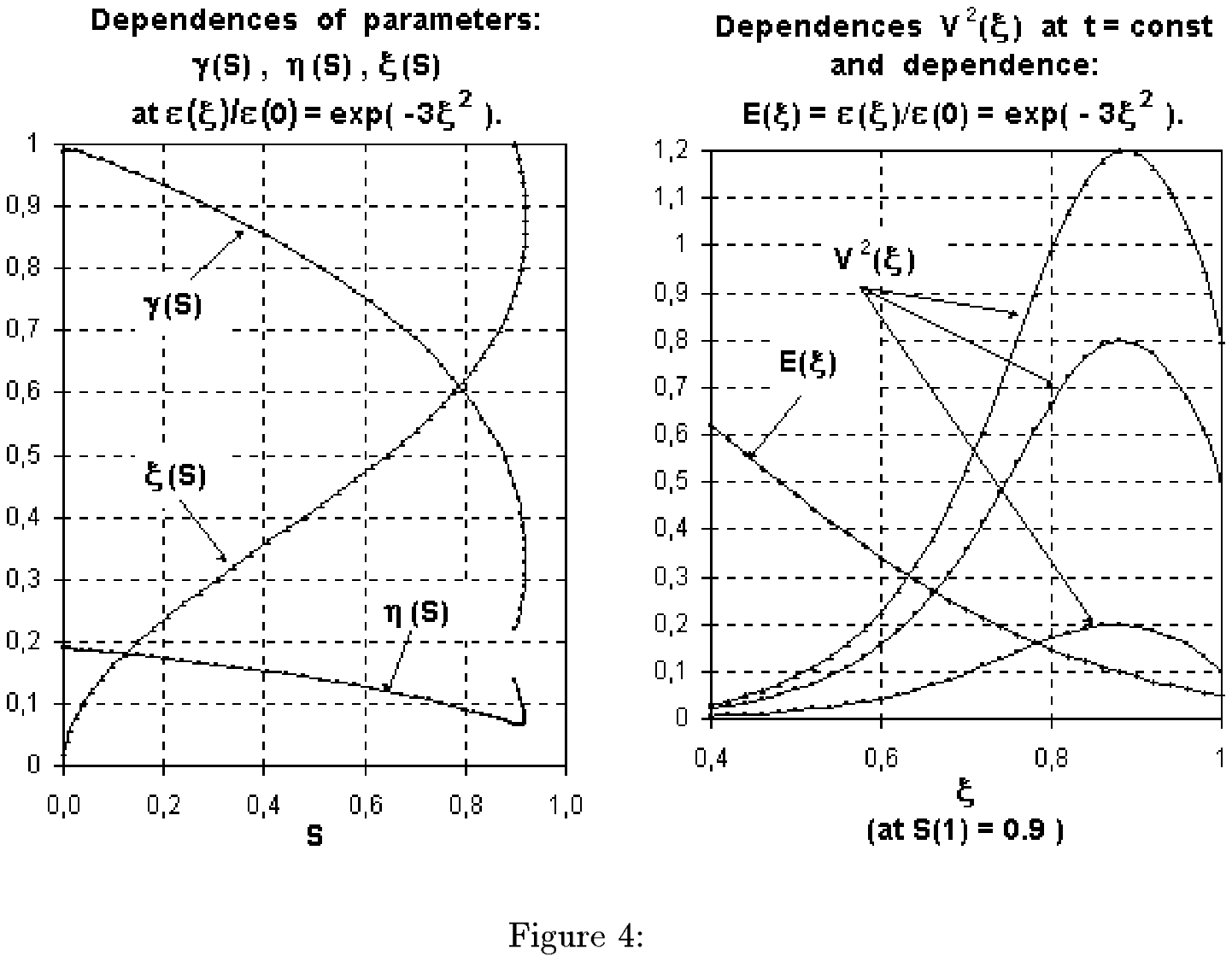}
\label{4}
\end{figure}

The numerical calculations confirm that the analytical
results of this Appendix remain valid for the real equation of
state of matter: 
$P = \alpha\varepsilon$.  Figure \ref{4} shows plots of $\gamma(S)$,
$\xi(S)$, and $\eta(S)$ for the special case of a Gaussian density
distribution: 
$\varepsilon_0(\xi)/\varepsilon_0(0) = \exp(-3\xi^2)$.  Comparing this figure
with Fig. \ref{2}, we can see that the $\gamma(S)$ curve in Fig.
\ref{4} crosses the $\gamma(S)$ curves in Fig. \ref{2} in the downward
direction roughly at the point $S \approx 0.92$, if we proceed
from $\xi=0$ to $\xi=1$.  As can be seen in Fig. \ref{4}, the point
$S \approx 0.92$ corresponds to $\xi \approx 0.85$ and $\eta \approx
0.1$; therefore, since the region where $V' > 0$ is located above
and to the right of the curves in Fig. \ref{2} and the region
where $V' < 0$ is located below and to the left of these curves,
the point $\xi \approx 0.85$ should be a local maximum on the
$V(\xi)$ curve for a constant value of $t$.

This analytical result is confirmed by a numerical
calculation of $V(\xi)|_{t={\rm const}}$ curves, whose results are shown
in Fig. \ref{5} with the predicted maxima.

To conclude this Appendix we would like to say a few words
regarding the initial characteristics and distribution of the
matter.

When the equations of the model were brought into
dimensionless form, it was found that the solution for a
spherical layer of matter with the index $R$ is completely
specified by three dimensionless parameters in the initial moment
in that layer: $0 < S < 1$, $0 < \gamma < 1$, and $0 < \eta < 1$.  This
corresponds to assigning the initial conditions for the
gravitational potential and two parameters which determine the
distribution of the matter and the pressure gradient near the
point under consideration.  Thus, upon integration we at once
find a whole family of self-similar solutions,\cite{8)}$^)$ which can be
characterized by these three parameters alone and which contains
the dependence on the other layers of matter above and below the
radius $R$ considered.

We thank N. S. Kardashev, V. L. Ginzburg, B. V. Komberg, V.
N. Lukash, and Yu. M. Bruk, as well as all the participants  in
the seminars of the Division of Theoretical Physics and the
Astrocosmic Center of the P. N. Lebedev Physics Institute of the
Russian Academy of Sciences for fruitful discussions of this work
and for their important comments.

With our sincerest gratitude we recall D. A. Kirzhnits, with
whom we formulated the ideas and initial approaches used in the
present work.

This work was supported by the Russian Foundation for Basic
Research (Grant No. 96-15-96616).

Translated by P. Shelnitz


\begin{thebibliography}{99}
\bibitem[2)]{2)}In case under discussion the word ``dust'' does
not, in any way, imply the absence of interactions between the
particles, except in the case of $\alpha = 0$.
\bibitem[3)]{3)}When there is an horizon, the components of the
metric in the reference frame which is stationary at infinity,
like the Schwarzchild metric, must acquire singularities as the
surface of that horizon is approached.
\bibitem[4)]{4)}Here we are dealing with the so-called apparent
horizon, or, as it is also called, trapping surface, whose
definition has been given in Refs. \cite{5} and
\cite{6}.  As for the event horizon, as will be seen below,
after all the matter falls, it coincides with the apparent
horizon (for further details, see Appendix 1).
\bibitem[6)]{6)}We consider just such a case, since intersection
of the layers would signify the appearance of an infinite density
already at a radius $r\neq0$, which is forbidden by cosmic
censorship (see, for example, Ref. \cite{6}).
\bibitem[7)]{7)}The significance of $h_{\rm cr}$ is that, according to
(3a), $\ddot{r}=0$ when $h=h_{\rm cr}$ in the initial moment, i.e.,
the system is in an unstable equilibrium.  Here and in the
following $\partial_\xi = \partial/\partial\xi$.
\end{thebibliography}
\end{document}